\documentclass[preprint2,10.5pt]{aastex}
\begin{document}

\title{\large{\rm{THE COLOR-MAGNITUDE DIAGRAM OF NGC 2264}}} 
\author{D. G. Turner}
\affil{Saint Mary's University, Halifax, Nova Scotia, Canada}
\email{turner@ap.smu.ca}

\begin{abstract}
Existing photometry for NGC 2264 tied to the Johnson \& Morgan (1953) {\it UBV} system is reexamined and, in the case of the original observations by Walker (1956), reanalyzed in order to generate a homogeneous data set for cluster stars. Color terms and a Balmer discontinuity effect in Walker's observations were detected and corrected, and the homogenized data were used in a new assessment of the cluster reddening, distance, and age. Average values of $E_{B-V}= 0.075 \pm0.003$ s.e. and $V_0${\it --M}$_V=9.45 \pm 0.03$ s.e. ($d = 777 \pm12$ pc) are obtained, in conjunction with an inferred cluster age of $\sim 5.5 \times 10^6$ yr from pre-main-sequence members and the location of the evolved, luminous, O7 V((f)) dwarf S Mon relative to the ZAMS. The cluster main sequence also contains gaps that may have a dynamical origin. The dust responsible for the initial reddening towards NGC 2264 is no more than 465 pc distant, and there are numerous, reddened and unreddened, late-type stars along the line of sight that are difficult to separate from cluster members by standard techniques, except for a small subset of stars on the far side of the cluster embedded in its gas and dust and background B-type ZAMS members of Mon OB2. A compilation of likely NGC 2264 members is presented. Only 3 of the 4 stars recently examined by asteroseismology appear to be likely cluster members. NGC 2264 is also noted to be a double cluster, which has not been mentioned previously in the literature.
\end{abstract}


\section{{\rm \footnotesize INTRODUCTION}}
``How often have I said to you that when you have eliminated the impossible, whatever remains, however improbable, must be the truth?'' Sherlock Holmes to Dr. Watson (Conan~Doyle 1890). 

In the study of open clusters a similar totalogy could be paraphrased as ``when you have eliminated the effects of extinction and differential reddening for cluster stars, the resulting cluster color-magnitude diagram, however unusual, must represent an accurate picture of the temperatures and luminosities of cluster stars.'' That philosophy was demonstrated to be the case for relatively young (10$^7$--10$^8$ yr old) open clusters by Turner (1996), as well as for the extremely young ($3.5 \times 10^6$ yr old) cluster IC 1590 (Guetter \& Turner 1997). The last study demonstrated how closely pre-main-sequence members of IC 1590 matched model isochrone predictions by Palla \& Stahler (1993) converted to the Johnson \& Morgan (1953) {\it UBV} system. 

The situation for other young clusters is more complicated, primarily because most of the original {\it UBV} studies of the brighter members of the class (NGC 2264, NGC 6530, IC 1546, and NGC 6611) were made by Merle Walker (Walker 1956, 1957, 1959, 1961, respectively) using a detector system that was not an ideal match to the Johnson system, although that was not realized at the time. For example, later studies of NGC 6611 (Hiltner \& Morgan 1969) and NGC 2264 (e.g., Mendoza \& G\'{o}mez 1980) noted differences between Walker's photometry and observations tied more closely to the {\it UBV} system. The origin of such differences can be explained by the work of Moffat \& Vogt (1977) and Guti\'{e}rrez-Moreno, Moreno \& Cort\'{e}z (1981), who noted that systematic errors, specifically in {\it U--B} measures, can arise from the manner in which the Balmer discontinuity in the continua of early-type stars is sampled by non-standard telescope/filter/detector systems, as well as by the treatment of atmospheric extinction (see Cousins \& Caldwell 2001). The differences in the case of NGC 6611 are fairly extreme, amounting to offsets of $0.10$ in {\it U--B} and $0.03$ in {\it B--V} (Hiltner \& Morgan 1969).

The purpose of the present study is to redo Walker's original study of NGC 2264 (Walker 1956) in order to generate a new reddening-free and extinction-corrected color-magnitude diagram for the cluster. The cluster has been studied many times previously, but never for the purpose of improving upon Walker's results. Recent detections of non-radial pulsation via asteroseismology in many of the pre-main-sequence members of NGC 2264 (e.g., Zwintz 2008; Kallinger, Zwintz \& Weiss 2008; Guenther et al. 2009) make it imperative to have a clear picture of the evolutionary status and exact H-R diagram location of cluster stars.

\section{{\rm \footnotesize ADJUSTING EXISTING {\it UBV} OBSERVATIONS}}
NGC 2264 has a rich history of observation, and has been studied fruitfully using many optical photometric systems. The emphasis here is on {\it UBV} photometry, which has certain advantages over other photometric systems for analyzing interstellar reddening; in most optical photometric systems interstellar reddening corrections are made using mean reddening laws, occasionally established incorrectly, that may not describe the reddening applicable to the cluster under study (e.g., Turner 1989, 1994, 1996). Str\"{o}mgren system photometry is similar enough to {\it UBV} photometry that it can be transformed accurately to the latter for most early-type stars (Turner 1990).

\begin{figure}[!t]
\begin{center}
\includegraphics[width=0.4\textwidth]{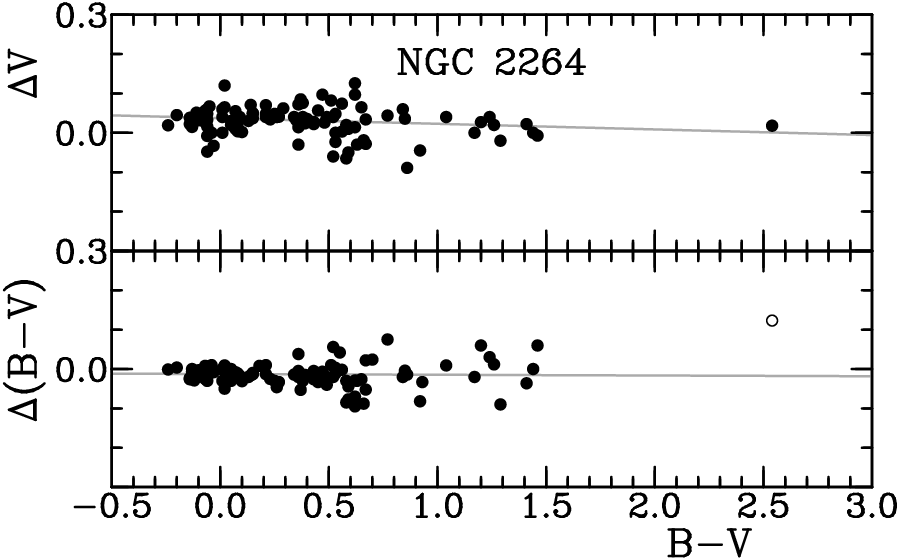}
\end{center}
\caption{\small{Differences, $\Delta V$ (upper) and $\Delta (B$--$V)$ (lower), between the derived mean {\it UBV} data from the literature and the Walker (1956) values, as a function of Walker's tabulated {\it B--V} color. The derived trends are indicated, and an open circle denotes the red star not used in the $\Delta (B-V)$ comparison.}}
\label{fig1}
\end{figure}

The procedure adopted here was to compare existing {\it UBV} observations, identified as being closely linked to the Johnson system and including transformations of published Str\"{o}mgren photometry for cluster stars, with the original measures of Walker (1956), and to adjust the latter for systematic effects prior to forming optimized values for individual stars. {\it UBV} observations not clearly tied closely to the Johnson system were not used, our experience being that systematic effects in such data are often non-linear, making them very difficult to treat reliably.

Sources of {\it UBV} photometry for NGC 2264 that contain data for sizable samples of stars are presented by Mendoza \& G\'{o}mez (1980), Per\'{e}z, Th\'{e} \& Westerlund (1987), and Kwon \& Lee (1983). Observations for more restricted samples can be found in papers by Johnson \& Morgan (1955), Hiltner (1956), Karlsson (1966), Turner (1976b), Macmillan (1977), Echevarr\'{i}a, Roth \& Warman (1979), Clar\'{i}a (1985), Oja (1991), and Zwintz (2008). Suitable Str\"{o}mgren system photometry that can be converted to equivalent {\it UBV} data, at least for early-type stars, is presented by Crawford, Barnes \& Golson (1971), Strom, Strom \& Yost (1971), Morrison (1975), Gronbech \& Olsen (1976), Perry \& Johnston (1982), Lindroos (1983), Olsen (1983), Per\'{e}z et al. (1988), Mendoza, Rolland \& Rodriguez (1990), Knude (1992), Handler (1999), Pe\~{n}a, Peniche \& Cervantes (2000), and Kalinger et al. (2008). The relationships derived by Turner (1990) were used to convert the latter observations to the Johnson system.

\begin{figure}[!t]
\begin{center}
\includegraphics[width=0.4\textwidth]{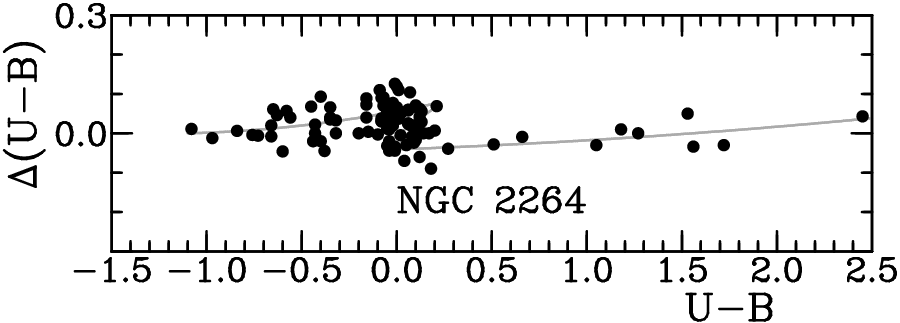}
\end{center}
\caption{\small{Differences $\Delta (U$--$B)$ between the derived mean {\it UBV} data from the literature and Walker's (1956) values, as a function of Walker's tabulated {\it U--B} color. The derived Z-shaped curve represents a least squares fit to three segments of the data.}}
\label{fig2}
\end{figure}

A comparison of the average {\it BV} values from such an analysis with Walker's (1956) photometry is presented in Fig.~\ref{fig1}. The differences in $\Delta V$ and $\Delta(B$--$V)$ as a function of {\it B--V} color are generally small, the best-fitting results being: $\Delta V=+0.037(\pm0.003) - 0.014(\pm0.006)$({\it B--V}) and $\Delta$({\it B--V}) $=-0.013(\pm0.002) - 0.002(\pm0.005)$({\it B--V}). There is a slight offset in color for Walker's observations (the reddest star being omitted for reasons of possible brightness variability) and the visual magnitudes display a reasonably distinct color dependence. The comparison of {\it U--B} colors in Fig.~\ref{fig2} is more interesting. There is increased scatter for stars with colors near {\it U}--$B \simeq 0.0$ that has the distinct signature of a Balmer discontinuity effect as described by Moffat \& Vogt (1977). The data were therefore matched to linear or quadratic functions in separate segments, with the resulting zig-zag relationship plotted in the figure adopted for further analysis. The Moffat \& Vogt (1977) results were used as a guide for interpreting the trends in the $\Delta(U$--$B)$ data.

\setcounter{table}{0}
\begin{table*}
\caption[]{Optimized {\it UBV} observations for NGC 2264 stars}
\label{tab1}
\begin{center}
\footnotesize
\begin{tabular*}{0.89\textwidth}{@{\extracolsep{-0mm}}ccccccccccccccccc}
\hline \noalign{\smallskip}
Star &{\it V} &{\it B--V} &{\it U--B} &n & &Star &{\it V} &{\it B--V} &{\it U--B} &n & &Star &{\it V} &{\it B--V} &{\it U--B} &n \\
\noalign{\smallskip} \hline \noalign{\smallskip}
1 &14.08 &0.91 &0.33 &3 & &80 &15.27 &1.52 &0.69 &1 & &159 &10.99 &0.06 &0.00 &4 \\
2 &9.72 &0.24 &0.18 &18 & &81 &16.29 &1.24 &0.04 &1 & &161 &14.99 &0.66 &--0.60 &1 \\
3 &14.60 &1.43 &1.12 &1 & &83 &7.95 &--0.16 &--0.83 &21 & &164 &13.32 &0.83 &0.21 &1 \\
4 &13.49 &0.42 &--0.03 &1 & &84 &11.99 &0.57 &0.07 &5 & &165 &10.99 &0.14 &0.11 &5 \\
5 &14.14 &0.53 &0.05 &3 & &85 &15.00 &1.08 &0.76 &1 & &168 &15.15 &1.24 &1.02 &1 \\
6 &8.21 &0.36 &--0.04 &6 & &87 &10.77 &0.21 &0.16 &3 & &169 &13.42 &0.77 &0.22 &1 \\
7 &7.79 &--0.12 &--0.60 &6 & &88 &9.05 &--0.12 &--0.66 &13 & &172 &10.09 &--0.06 &--0.40 &2 \\
8 &13.64 &1.25 &0.86 &4 & &89 &16.36 &1.04 &--0.57 &1 & &173 &16.43 &1.27 &0.75 &1 \\
9 &11.99 &2.02 &2.18 &2 & &90 &12.67 &0.19 &--0.04 &16 & &175 &16.14 &1.39 &0.45 &1 \\
10 &11.26 &0.56 &0.19 &4 & &91 &12.33 &0.64 &0.00 &3 & &176 &15.80 &1.57 &0.83 &1 \\
11 &16.06 &0.86 &0.03 &1 & &92 &11.66 &0.85 &0.38 &6 & &177 &9.08:: &0.74:: &0.05 &4 \\
12 &15.49 &1.28 &0.15 &1 & &93 &13.21 &0.87 &0.32 &2 & &177B &11.98 &0.57 &0.00 &1 \\
13 &12.05 &1.66 &1.46 &2 & &94 &10.45 &0.41 &--0.04 &2 & &178 &7.18 &--0.20 &--0.98 &13 \\
14 &14.85 &0.79 &0.17 &1 & &95 &16.07 &1.10 &--0.05 &1 & &178B &10.23 &0.16 &--0.12 &2 \\
15 &14.66 &1.20 &0.65 &1 & &96 &14.10 &1.09 &1.22 &1 & &179 &9.96 &0.00 &--0.18 &5 \\
16 &16.12 &0.95 &0.33 &3 & &98 &11.77 &0.57 &0.00 &2 & &180 &12.92 &0.51 &--0.08 &2 \\
17 &12.89 &0.37 &0.23 &4 & &99 &10.86 &0.40 &--0.03 &5 & &181 &10.06 &--0.05 &--0.30 &3 \\
18 &15.23 &0.84 &0.20 &2 & &100 &10.04 &0.12 &0.09 &10 & &182 &10.33 &0.06 &0.08 &4 \\
19 &15.54 &0.82 &--0.06 &1 & &103 &10.08 &0.02 &--0.10 &2 & &183 &15.24 &0.96 &0.63 &1 \\
20 &10.30 &0.42 &0.16 &26 & &104 &11.40 &0.22 &0.13 &7 & &184 &14.16 &0.99 &--0.05 &1 \\
21 &14.09 &0.77 &0.22 &2 & &105 &15.13 &1.06 &--0.33 &1 & &186 &15.63 &1.54 &0.92 &1 \\
22 &17.08 &1.07 &0.16 &3 & &106 &13.31 &0.71 &0.38 &1 & &187 &9.25 &--0.08 &--0.33 &5 \\
23 &13.75 &0.82 &0.49 &2 & &107 &8.86 &--0.08 &--0.46 &15 & &189 &11.36 &0.54 &0.01 &4 \\
24 &8.57 &--0.06 &--0.31 &5 & &108 &12.05 &0.58 &0.07 &6 & &190 &12.29 &0.67 &--0.01 &1 \\
25 &7.83 &0.37 &0.00 &11 & &109 &9.12 &--0.11 &--0.53 &5 & &193 &9.81 &0.23 &0.12 &6 \\
26 &11.78 &0.47 &0.01 &3 & &112 &10.81 &--0.01 &--0.11 &6 & &195 &12.66 &0.52 &--0.05 &4 \\
27 &12.08 &0.52 &0.21 &3 & &114 &11.55 &0.52 &0.07 &4 & &196 &11.72 &0.52 &--0.03 &3 \\
28 &12.34 &0.46 &--0.08 &4 & &115 &14.44 &1.00 &0.37 &1 & &197 &16.30 &0.92 &--0.08 &1 \\
29 &10.16 &0.43 &0.00 &5 & &116 &11.66 &0.54 &0.09 &5 & &202 &9.02 &0.07 &--0.59 &10 \\
30 &10.78 &0.03 &--0.02 &6 & &117 &13.55 &0.69 &0.25 &1 & &203 &12.93 &0.75 &0.14 &1 \\
31 &10.56 &0.35 &0.03 &5 & &118 &11.80 &0.55 &--0.04 &2 & &205 &10.64 &0.33 &--0.02 &5 \\
32 &13.02 &0.76 &0.05 &1 & &121 &12.14 &0.42 &--0.27 &1 & &206 &9.50 &0.09 &--0.34 &5 \\
33 &11.68 &2.63 &2.49 &5 & &125 &12.29 &0.61 &0.02 &2 & &209 &11.37 &0.37 &0.02 &4 \\
34 &10.94 &0.40 &0.05 &1 & &127 &15.77 &1.48 &0.85 &1 & &210 &13.30 &0.76 &0.09 &1 \\
35 &10.34 &0.08 &0.06 &4 & &128 &11.03 &0.26 &0.10 &3 & &212 &7.51 &--0.16 &--0.75 &20 \\
36 &11.01 &0.01 &--0.01 &5 & &131 &4.64 &--0.24 &--1.07 &63 & &215 &9.32 &0.07 &--0.14 &10 \\
37 &8.08 &1.49 &1.66 &7 & &132 &10.22 &--0.04 &--0.29 &6 & &216 &11.72 &0.76 &0.14 &1 \\
38 &10.98 &1.04 &0.64 &2 & &133 &13.85 &1.05 &0.61 &1 & &217 &13.60 &0.81 &--0.34 &2 \\
39 &11.35 &0.11 &0.11 &2 & &134 &12.42 &0.82 &0.01 &3 & &220 &9.72 &0.46 &--0.06 &3 \\
43 &10.57 &0.20 &0.16 &8 & &136 &15.17 &1.56 &1.05 &1 & &221 &12.15 &0.39 &--0.07 &4 \\
46 &9.23 &0.21 &0.18 &14 & &137 &9.93 &--0.08 &--0.40 &5 & &222 &9.92 &0.14 &0.18 &6 \\
50 &8.15 &--0.15 &--0.74 &19 & &138 &10.19 &0.06 &--0.01 &5 & &223 &10.93 &0.33 &0.03 &5 \\
54 &14.27 &1.23 &1.01 &1 & &139 &13.33 &1.24 &0.81 &2 & &224 &11.50 &0.52 &0.15 &2 \\
58 &15.52 &0.96 &0.60 &1 & &141 &14.72 &1.14 &0.85 &1 & &225 &13.22 &0.57 &--0.01 &2 \\
60 &12.49 &1.01 &0.67 &2 & &142 &8.86 &--0.09 &--0.53 &10 & &226 &9.63 &0.14 &0.16 &7 \\
62 &12.37 &0.50 &--0.18 &2 & &143 &10.63 &0.06 &--0.10 &1 & &227 &11.80 &0.52 &--0.06 &1 \\
64 &15.36 &0.93 &0.39 &1 & &144 &13.85 &1.36 &0.30 &1 & &228 &11.13 &0.35 &0.01 &5 \\
65 &11.76 &0.49 &--0.04 &2 & &145 &10.67 &0.04 &0.02 &6 & &229 &8.54 &1.23 &1.03 &9 \\
66 &12.43 &0.71 &--0.14 &4 & &146 &14.48 &1.08 &0.75 &1 & &230 &12.36 &1.31 &0.52 &1 \\
67 &10.88 &0.59 &--0.41 &5 & &147 &11.00 &0.80 &0.38 &2 & &231 &8.99 &--0.14 &--0.64 &7 \\
68 &11.75 &0.68 &0.11 &7 & &148 &13.63 &0.68 &0.05 &1 & &232 &9.83 &0.00 &--0.02 &5 \\
69 &8.28 &1.38 &1.54 &10 & &150 &14.13 &0.98 &1.00 &2 & &233 &9.57 &0.56 &0.05 &3 \\
70 &11.17 &0.57 &0.06 &5 & &151 &12.56 &0.50 &--0.01 &2 & &234 &12.43 &0.46 &--0.05 &2 \\
72 &12.36 &0.50 &--0.10 &1 & &152 &9.15 &--0.07 &--0.41 &14 & &235 &13.47 &0.51 &0.38 &3 \\
73 &9.35 &0.84 &0.48 &8 & &153 &15.89 &1.22 &0.33 &1 & &236 &11.39 &0.63 &0.12 &5 \\
74 &6.88 &--0.11 &--0.63 &16 & &154 &12.62 &0.79 &0.03 &2 & &237 &9.45 &1.43 &1.26 &5 \\
76 &14.14 &0.75 &0.38 &1 & &155 &16.50 &1.41 &0.96 &1 & &238 &9.97 &1.25 &1.56 &3 \\
77 &14.53 &1.17 &0.90 &1 & &156 &14.68 &1.23 &0.66 &1 & &239 &9.34 &1.15 &1.02 &4 \\
78 &15.42 &1.21 &--0.39 &1 & &157 &10.04 &--0.03 &--0.31 &6 & & & & & & \\
79 &15.90 &0.55 &--1.21 &1 & &158 &10.35 &0.34 &0.07 &6 & & & & & & \\
\noalign{\smallskip} \hline
\end{tabular*}
\end{center}
$\;\;\;\;\;\;\;\;\;\;\;$\small{Star numbers from Walker (1956) and WEBDA, except as noted in text.}
\end{table*}

\begin{figure}[!t]
\begin{center}
\includegraphics[width=0.40\textwidth]{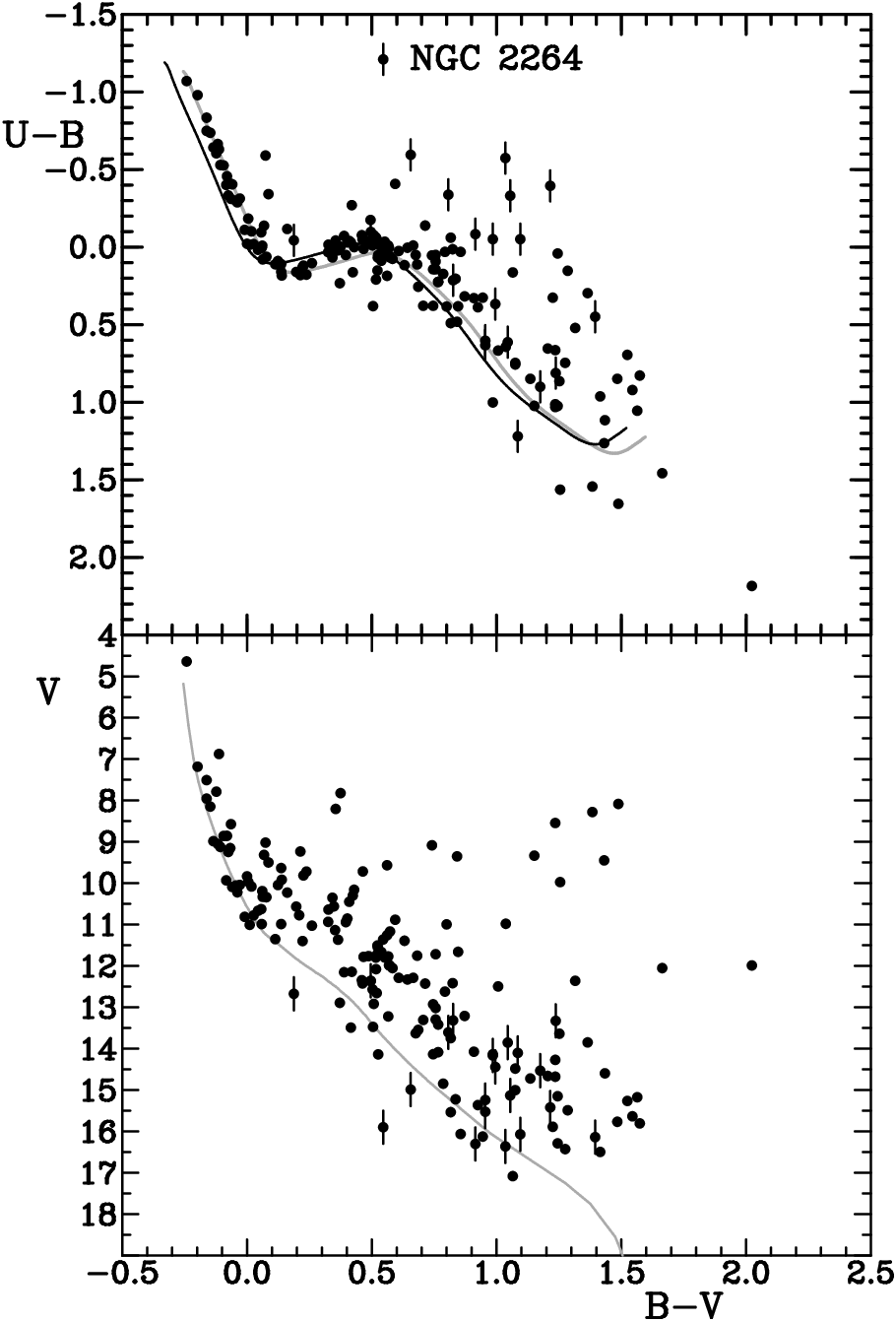}
\end{center}
\caption{\small{{\it UBV} color-color (upper) and color-magnitude (lower) diagrams for NGC 2264 stars with photoelectric observations. The intrinsic color-color relation is the black curve, while gray curves denote that relation and the zero-age main sequence (ZAMS) reddened by $E_{B-V} = 0.075$ and {\it V--M}$_V = 9.68$. Stars exhibiting emission at $H\alpha$ are denoted by vertical lines.}}
\label{fig3}
\end{figure}

The power of the present approach to homogenizing existing {\it UBV} photometry for NGC 2264 can be seen in the results. Walker's (1956) {\it UBV} observations were adjusted with the relations plotted in Figs.~\ref{fig1}~and~\ref{fig2} and averaged with the other data, weighted by the number of observations for each observer, to form optimized values, summarized in Table~\ref{tab1}. The resulting data are plotted in color-color and color-magnitude diagrams in Fig.~\ref{fig3}, and display a variety of traits that differ in subtle ways from the original versions published by Walker (1956). As in Walker's study, stars exhibiting $H\alpha$ emission (Herbig 1954) are identified.

First, the sequence of reddened B-type stars closely fits the intrinsic relation for dwarfs with $E_{B-V}=0.075$, similar to what is seen in Stock 16 (Turner 1985, 1996) where there is no evidence for differential reddening. There is sufficient scatter, however, to suggest a better match to NGC 2422 (Turner 1996), which displays a color excess spread of $\Delta E_{B-V}= 0.05$. Second, the colors of stars near the A0-star kink fit the reddened relation closely, much better than for Walker's (1956) data. Apparently the correction for the Balmer discontinuity effect adopted in Fig.~\ref{fig2} offers a satisfactory solution to a serious systematic error in the original Walker  photometry. Third, there are a number of FGK stars in the field with the colors of unreddened field stars, suggesting a possible source of contamination for the cluster color-magnitude diagram. There is also little evidence to indicate that heavily reddened stars are cluster members, given that the data do not display a continuous run of reddening, as in NGC 1647 (Turner 1992, 1996). It can be further noted that there are a number of objects with colors that are inconsistent with those of stars of normal reddening. Such an effect may be the result of extreme youth or contamination of the photometry by bright surrounding nebulosity. Many display $H\alpha$ emission.

The nature of the interstellar reddening in the direction of NGC 2264 was studied previously by Turner (1976b), yielding a {\it UBV} reddening slope of $E_{U-B}/E_{B-V} = 0.77$ and a ratio of total-to-selective extinction for early-type stars in the region of $R = A_V/E_{B-V} = 3.2 \pm 0.2$. Those values were adopted here in an analysis of the {\it UBV} data of Table~\ref{tab1}, and the resulting color excesses $E_{B-V}$ for individual stars were corrected for the color dependence caused by bandwidth effects in the {\it UBV} system (Schmidt-Kaler 1961; Fernie 1963; Buser 1978).

The field of NGC 2264 contains just two groups of OB stars. One displays only a small amount of interstellar reddening and consists of massive members of the cluster. The other consists of more heavily reddened ZAMS members of Mon OB2 lying $\sim 800$ pc beyond NGC 2264. The reddening separation of AFGK-type stars in terms of distance is less clear cut, as discussed in \S4.

The numbering of stars in Table~\ref{tab1} is generally that of Walker (1956), although with two additions and a few corrections. In the study by Pe\~{n}a et al. (2000), for example, some stars are misidentified. Their observation for star 14 is a second measurement for star 24, that for star 31 is for star 26, the two stars identified as 77A and 77B are measures for star 177 of Walker (1956) with the A and B notation reversed, the data for star 106 are for star 112, and the observation for star 169 is inconsistent with its measured brightness by Walker (1956) or with those of any bright stars in the cluster. It has therefore been omitted from the analysis. The observations by Echevarr\'{i}a et al. (1979) for the multiple system surrounding S Mon (ADS 5322, star 131) include measures for a star, ADS 5322Ad, that is star 121 of Walker (1956), but with a much more reasonable brightness than that obtained by Johnson for Walker's study. It has been included in Table~\ref{tab1} with Johnson's measures omitted. Likewise, the data from Lindroos (1983) include separate measures for a bright companion of Walker star 178, HD 47887B, so it has also been included.

\section{{\rm \footnotesize PARAMETERS FOR NGC 2264}}
\begin{figure}[!t]
\begin{center}
\includegraphics[width=0.4\textwidth]{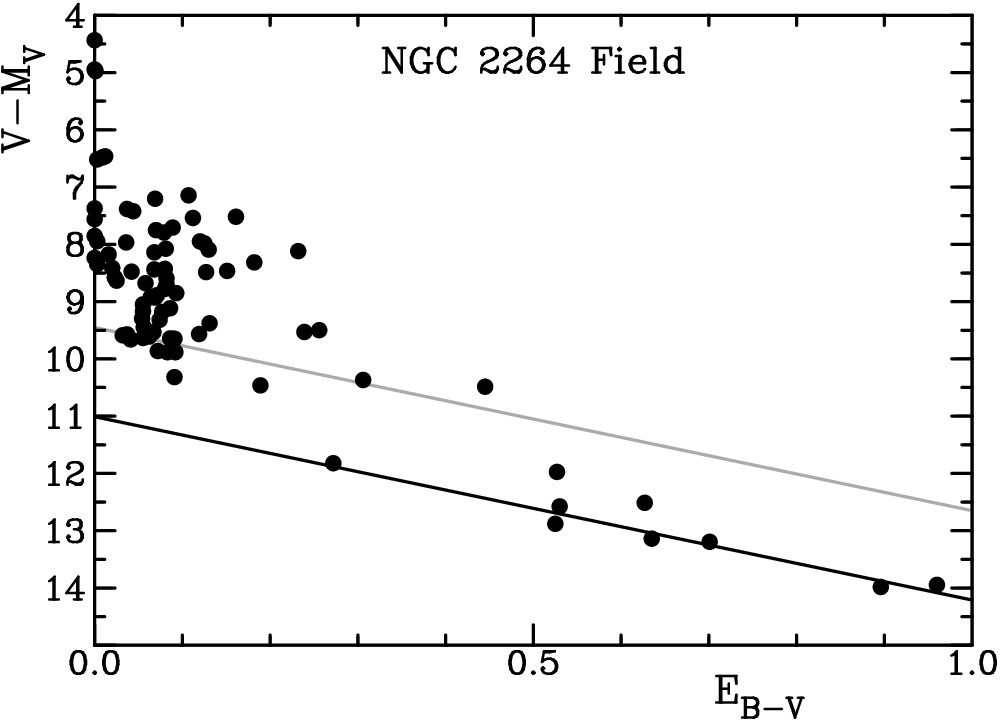}
\end{center}
\caption{\small{Variable extinction diagram for early-type stars in NGC 2264, for ZAMS values of $M_V$. The sloped lines represent $R=3.2$ for a ZAMS fit to NGC 2264 (gray line) and for Mon OB2 (black line), the last from Turner (1976b).}}
\label{fig4}
\end{figure}

Since the scatter of reddened B-type stars in NGC 2264 seen in Fig.~\ref{fig3} is larger than the uncertainties in the observed colors of at most $\pm 0.01$, it is appropriate to analyze the data of Table~\ref{tab1} using the variable-extinction method (see Turner 1976a). A reddening line of slope $E_{U-B}/E_{B-V}=0.77$ (Turner 1976b) was therefore applied to the observations, with larger slopes being used for G, K, and M-type stars, and the data were dereddened to the intrinsic relation for dwarfs (see Turner 1996). Absolute magnitudes appropriate for a zero-age main-sequence (ZAMS) star were assigned to each analyzed star on the basis of its derived intrinsic color, where the ZAMS values are from Turner (1976a, 1979). The results are displayed in Fig.~\ref{fig4}, which reveals much about the extinction in the direction of NGC 2264.

An initial inference from Fig.~\ref{fig4} is that there are unreddened stars towards NGC 2264 with intrinsic distance moduli $V_0$--$M_V$ ranging between 4 and 8.34, which indicates that the dust clouds responsible for the reddening of most stars along the line of sight must lie at distances of $\sim 465$ pc, a result consistent with the findings of Neckel \& Klare (1980) for surrounding fields. The fact that most cluster stars lie in regions of bright nebulosity, both emission and reflection, and yet have only small reddenings, indicates that the cluster must lie mostly in front of the gas and dust clouds that dominate the region. A few cluster stars have slightly larger reddenings that suggest they lie on the far side of the cluster and are imbedded in the dust clouds.

More heavily reddened cluster stars, on the other hand, display a remarkably close coincidence ($d = 1.6$ kpc, $R = 3.2$) with the spectroscopically-derived variable-extinction results of Turner (1976b) for OB stars in Mon OB2, implying that one can readily detect background objects lying on the far side of the cluster and beyond the associated gas and dust clouds. A detailed analysis of the full data set (\S4) also indicates that there are many other non-members of the cluster lying in the field, a few at about the same distance as the main group of cluster stars and many more foreground to the cluster but beyond the dust clouds producing most of the foreground reddening. A similar result can be inferred from the proper motion analysis of Vasilevskis, Sanders \& Balz (1965) for bright stars in the cluster field.

The distance to NGC 2264 can be obtained from the 13 B-type ZAMS stars in Fig.~\ref{fig4} that form a lower envelope to the main body of data for cluster stars. The resulting intrinsic distance modulus is $V_0-M_V=9.45 \pm 0.03$ s.e., corresponding to a distance of $777 \pm12$ pc. An almost identical distance was obtained by Turner (1976b) from an independent analysis of Walker's original observations in conjunction with spectroscopic data for members of Mon OB1 lying near the cluster. The average reddening for stars belonging to the main body of the cluster is $E_{B-V}=0.075 \pm0.003$ s.e., as noted earlier, with an observed dispersion from the standard deviation ($\sigma$) of $\Delta E_{B-V} = 0.06$, slightly larger than the reddening dispersion observed in NGC 2422, discussed above.  

The identification of likely cluster members is more complicated than the case for more distant groups, primarily because the cluster, because of its relative closeness, is spread out on the sky. The general appearance of NGC 2264 is also that of a double cluster, with no guarantee that the northern and southern components are of the same age and distance (although it does seem likely). For bright stars there are radial velocity measures and proper motion membership probabilities (Vasilevskis et al. 1965) for consideration, although neither is fully reliable, particularly in a direction roughly towards the Galactic anticenter, as is the case for NGC 2264 ($l = 203^{\circ}$). The radial velocities for individual stars summarized in WEBDA display large scatter for several stars, the signature of spectroscopic binaries, are biased towards bright cluster stars, and have similar values both for {\it bona fide} cluster members and nearby field stars, while the proper motions of stars across the line of sight at distances of 500--800 pc may be fairly similar. Compounding the situation is the apparent presence of a sizable population of AF-type stars in the immediate foreground to the cluster.

Likely cluster members were therefore identified in the present study by considering all available data for each star, in conjunction with the results for other cluster stars analyzed. The dereddening procedure from Fig.~\ref{fig3} is clearly not straightforward, given the potential non-physical solutions that arise for many of the faint red stars, many of which are variable or display $H\alpha$ emission. Initially it was believed that many of the latter objects might be cluster members, and dereddening solutions were adopted for them using the mean cluster reddening. However, the resulting distribution of pre-main-sequence stars in the cluster color-magnitude diagram differed from that for non-variable pre-main-sequence stars without $H\alpha$ emission, suggesting that they are either field stars or embedded objects that must be treated separately.

\begin{figure}[!t]
\begin{center}
\includegraphics[width=0.4\textwidth]{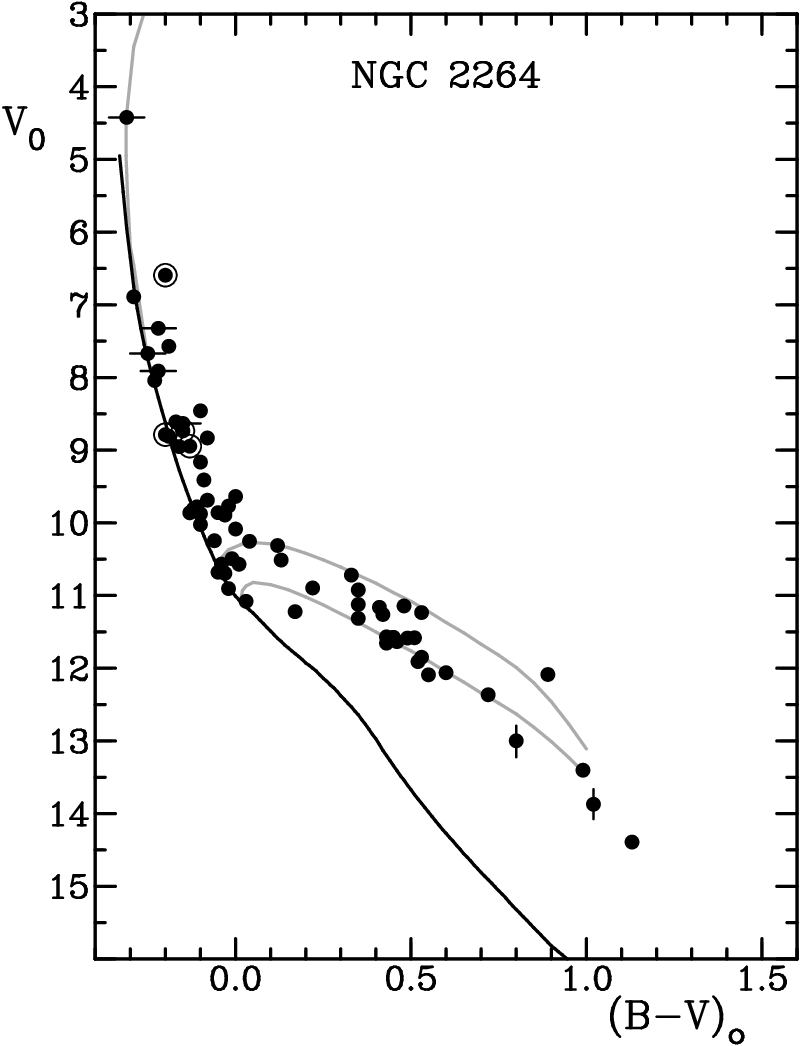}
\end{center}
\caption{\small{Reddening and extinction corrected color-magnitude diagram for likely members of NGC 2264 from the present study. The black curve represents the ZAMS for the derived cluster distance, and gray curves at the bottom are pre-main-sequence isochrones for $\log t = 6.5$ (upper) and 6.75 (lower). The gray curve at top is a post-main-sequence isochrone for $\log t = 6.7$. Known and suspected spectroscopic binaries are indicated by horizontal lines, rapidly-rotating stars are circled, and stars exhibiting emission at $H\alpha$ are denoted by vertical lines.}}
\label{fig5}
\end{figure}

A compilation of likely members of NGC 2264 from the present study is given in Table~\ref{tab2}, and the resulting reddening-free and extinction-free color-magnitude diagram for them is shown in Fig.~\ref{fig5}. The close fit of the 13 stars discussed above to the ZAMS is evident here, as are a few other noteworthy features. Known or suspected spectroscopic binaries are identified, and the bias towards bright cluster stars is evident. Stars identified spectroscopically as rapid rotators, from ``n'' or ``nn'' classifications or large $V \sin i$ values (Vogel \& Kuhi 1981), are also noted, as are the two stars displaying emission at $H\alpha$ (Walker 96, 133).

The identification of rapid rotators relates to the cluster main sequence, which appears to contain gaps, a feature seen in several other clusters (Mermilliod 1982) and attributed by Turner (1996) to close binary mergers and the resulting creation of rapidly-rotating merger products, which results in a color spread for cluster stars more luminous than each gap. That characteristic is evident for NGC 2264, as it is for the clusters discussed by Turner (1996). Most notable in the case of NGC 2264 is the B3 Vn star 74, the second most luminous likely cluster member, which lies above a main-sequence gap near $V_0\simeq7.2$, and the group of rapidly-rotating stars lying above the main-sequence gap near $V_0\simeq9.5$.

\setcounter{table}{1}
\begin{table}[!t]
\caption[]{Likely members of NGC 2264 from this study}
\label{tab2}
\begin{center}
\footnotesize
\begin{tabular*}{0.45\textwidth}{@{\extracolsep{-2.5mm}}ccccccccc}
\hline \noalign{\smallskip}
Star &$(B-V)_0$ &$E_{B-V}$ &$V_0$ & &Star &$(B-V)_0$ &$E_{B-V}$ &$V_0$ \\
\noalign{\smallskip} \hline \noalign{\smallskip}
7 &--0.19 &0.07 &7.57 & &128 &+0.22 &0.04 &10.89 \\
15 &+1.13 &0.08 &14.39 & &131 &--0.31 &0.07 &4.42 \\
24 &--0.10 &0.04 &8.46 & &132 &--0.10 &0.06 &10.02 \\
26 &+0.43 &0.04 &11.66 & &133 &+0.80 &0.27 &13.00 \\
30 &--0.04 &0.07 &10.57 & &137 &--0.12 &0.04 &9.81 \\
34 &+0.33 &0.07 &10.72 & &138 &--0.03 &0.09 &9.89 \\
35 &+0.00 &0.08 &10.08 & &142 &--0.17 &0.08 &8.61 \\
36 &--0.02 &0.03 &10.90 & &143 &--0.06 &0.12 &10.24 \\
39 &+0.03 &0.09 &11.08 & &145 &--0.01 &0.05 &10.49 \\
43 &+0.12 &0.08 &10.31 & &152 &--0.13 &0.06 &8.95 \\
50 &--0.22 &0.07 &7.91 & &154 &+0.72 &0.08 &12.37 \\
54 &+0.99 &0.27 &13.40 & &157 &--0.11 &0.08 &9.78 \\
60 &+0.89 &0.13 &12.09 & &159 &--0.03 &0.09 &10.69 \\
65 &+0.43 &0.06 &11.57 & &165 &+0.01 &0.13 &10.57 \\
68 &+0.53 &0.16 &11.23 & &172 &--0.13 &0.07 &9.86 \\
74 &--0.20 &0.09 &6.59 & &177B &+0.53 &0.04 &11.85 \\
83 &--0.25 &0.09 &7.67 & &178 &--0.29 &0.09 &6.89 \\
84 &+0.45 &0.13 &11.58 & &178B &--0.09 &0.26 &9.41 \\
87 &+0.13 &0.08 &10.51 & &179 &--0.08 &0.09 &9.69 \\
88 &--0.20 &0.08 &8.79 & &181 &--0.10 &0.06 &9.88 \\
91 &+0.52 &0.13 &11.91 & &182 &+0.04 &0.02 &10.26 \\
96 &+1.02 &0.07 &13.87 & &187 &--0.10 &0.03 &9.16 \\
98 &+0.51 &0.06 &11.58 & &189 &+0.48 &0.07 &11.14 \\
100 &+0.00 &0.13 &9.64 & &190 &+0.60 &0.07 &12.06 \\
103 &--0.05 &0.07 &9.86 & &202 &--0.23 &0.31 &8.04 \\
104 &+0.17 &0.06 &11.22 & &206 &--0.15 &0.24 &8.74 \\
107 &--0.15 &0.07 &8.63 & &209 &+0.35 &0.02 &11.31 \\
108 &+0.46 &0.13 &11.63 & &212 &--0.22 &0.06 &7.32 \\
109 &--0.16 &0.05 &8.95 & &215 &--0.08 &0.15 &8.83 \\
112 &--0.05 &0.04 &10.68 & &224 &+0.35 &0.18 &10.92 \\
114 &+0.41 &0.12 &11.16 & &227 &+0.45 &0.07 &11.58 \\
116 &+0.42 &0.13 &11.26 & &228 &+0.35 &0.00 &11.12 \\
118 &+0.49 &0.07 &11.59 & &231 &--0.19 &0.06 &8.81 \\
125 &+0.55 &0.06 &12.09 & &232 &--0.02 &0.02 &9.77 \\
\noalign{\smallskip} \hline
\end{tabular*}
\end{center}
\end{table}

Isochrones are also plotted in Fig.~\ref{fig5} for $\log t = 6.5$ and 6.75, the latter of which provides a good fit to the lower envelope of pre-main-sequence stars. An age of $\log t \simeq 6.7$ is also inferred for the most luminous cluster star, Walker 131 = S Mon = HD 47839 [(O7 V((f)), Walborn 1972], since it appears to have evolved away from the ZAMS. As in Guetter \& Turner (1997), the pre-main-sequence isochrones are adapted from the results of Palla \& Stahler (1994), while the post-main-sequence isochrone in Fig.~\ref{fig5} is from Meynet, Mermilliod \& Maeder (1993). Likewise, much like the case for IC 1590 (Guetter \& Turner 1997), it is not clear that the dispersion in luminosities for pre-main-sequence stars represents a spread in formation times. Binarity is also a possibility, particularly given the similarity in implied ages between lower envelope pre-main-sequence stars and S Mon. The implied age for members of NGC 2264 is therefore $\sim 5.5 \times 10^6$ yr.

\section{{\rm \footnotesize FIELD STARS IN NGC 2264}}
\begin{figure}[!t]
\begin{center}
\includegraphics[width=0.4\textwidth]{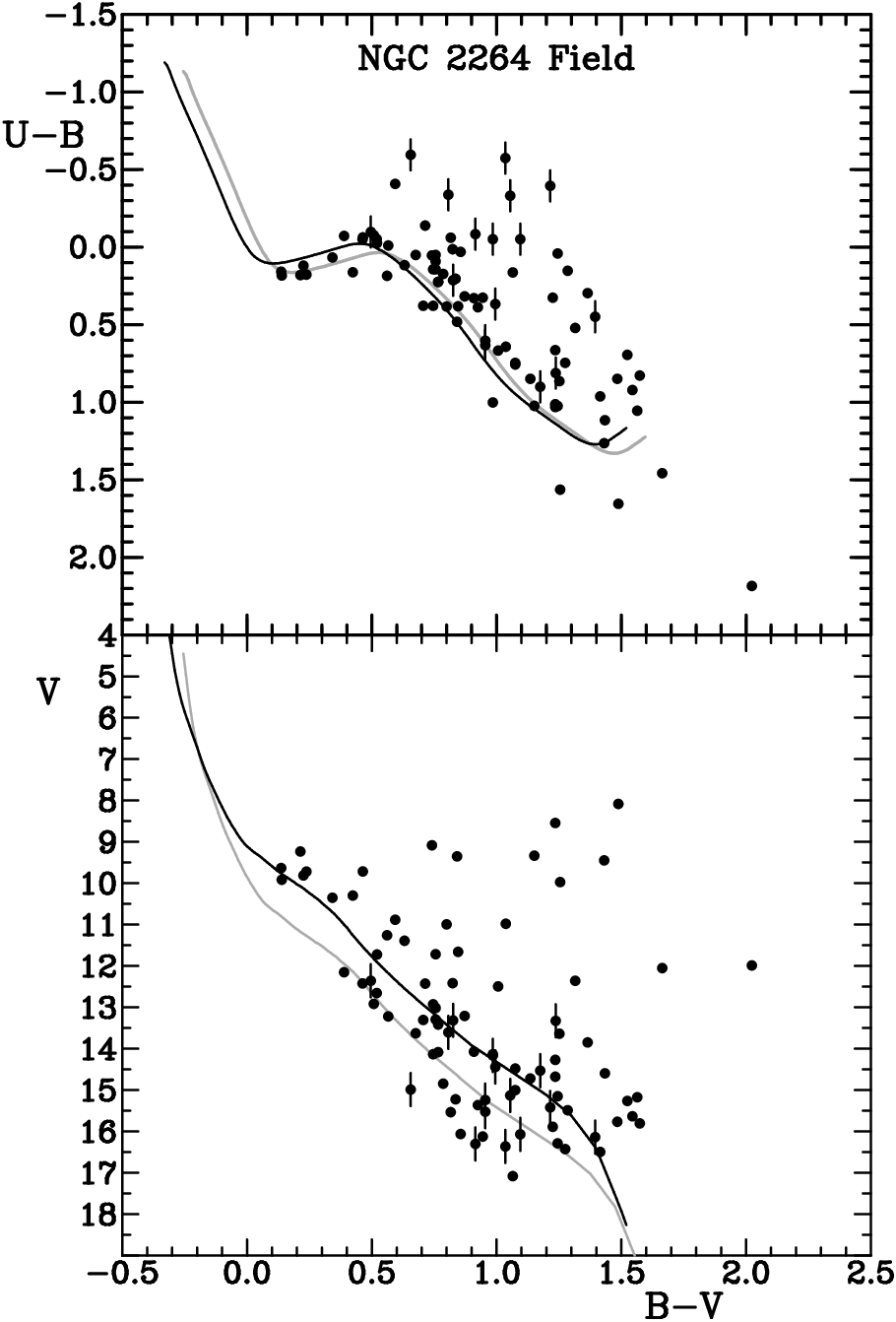}
\end{center}
\caption{\small{{\it UBV} color-color (upper) and color-magnitude (lower) diagrams for NGC 2264 stars with photoelectric observations. The intrinsic color-color and zero-age main sequence (ZAMS) relations are the black curves, the latter for $V_0$--$M_V = 7.55$, while gray curves denote the color-color relation and the ZAMS reddened by $E_{B-V} = 0.074$ and {\it V--M}$_V = 9.0$. Stars exhibiting emission at $H\alpha$ are denoted by vertical lines.}}
\label{fig6}
\end{figure}

Most previous studies of NGC 2264 have assumed that the AFGK-type stars lying above the main sequence in Walker's (1956) original cluster color-magnitude diagram are pre-main-sequence objects, with variability, $H\alpha$ emission, and large membership probabilities (Vasilevskis et al. 1965) for the same objects supplying confirmatory evidence. Yet when most such stars are assumed to be cluster members with reddenings similar to the cluster average, the resulting sequence of assumed pre-main-sequence objects runs parallel to the ZAMS, which is inconsistent with model isochrones (e.g., Palla \& Stahler 1993). That can be seen in Fig.~\ref{fig6}, which presents color-color and color-magnitude diagrams for stars from Table~\ref{tab1} that were rejected from cluster membership in Table~\ref{tab2}. Note that most of the stars displaying $H\alpha$ emission (Herbig 1954) fall in the plots of Fig.~\ref{fig6}.

A large number of the stars fall along an unreddened ZAMS for $V_0-M_V=7.55$ ($d=324$ pc), which suggests the presence of an intermediate-age population in this direction lying $\sim 150$ pc foreground to the dust clouds responsible for the initial interstellar extinction in NGC 2264. A smaller selection of stars fit the ZAMS reddened by $E_{B-V}=0.075$ for {\it V--M}$_V = 9.0$, corresponding to a distance of $570$ pc, $\sim 100$ pc beyond the foreground dust complexes but well foreground to NGC 2264. There also appear to be a few GK-type stars in Fig.~\ref{fig6} with distance moduli similar to that of NGC 2264. They probably lie at distances comparable to that of the cluster or slightly foreground to it, similar to the case for some of the stars in Fig.~\ref{fig4}. In this alternate interpretation of the observations, the region of NGC 2264 contains a sizable sample of stars belonging to the general field, the exceptions being the slightly reddened OB stars and many of the AF-type stars. $H\alpha$ emission does not appear to correlate with any particular status for individual stars.

The above discussion is relevant for recent observations of small-amplitude variable stars in NGC 2264 studied by means of asteroseimology. In the study made by Zwintz et al. (2009), for example, their star V1 seems likely to be a foreground star, stars V3 and V4 are possible pre-main-sequence objects, and star V2 is Walker 39, a newly-arrived ZAMS member of NGC 2264. Zwintz et al. (2009) refer to it as a pre-main-sequence $\delta$ Scuti variable; for that to be the case the star must be just on the verge of initiating core hydrogen burning. Such distinctions are important for establishing the pulsational characteristics of other targets for asteroseismology chosen according to their true location in the cluster color-magnitude diagram.

\section{{\rm \footnotesize DISCUSSION}}
One advantage of removing the effects of interstellar reddening and extinction from cluster color-magnitude diagrams is that the resulting distribution of data points must be closely linked to the true variations in effective temperature and luminosity for cluster members. In the case of NGC 2264, the color-magnitude diagram presented here in Fig.~\ref{fig5} provides a number of important insights into the evolutionary status of cluster stars. The main sequence gaps, for example, imply an advanced dynamical state, provided they arise from close binary mergers, as argued by Turner (1996). And the close coincidence in the cluster age inferred from both pre-main-sequence stars and the slightly-evolved S Mon imply that the formation of cluster stars was not spread out greatly in time, the creation of cluster stars probably consuming no more than a few hundred thousand years.

\subsection*{{\rm \scriptsize ACKNOWLEDGEMENTS}}
\scriptsize{The present study has used the open cluster data compilation maintained by WEBDA, operated at the Institute for Astronomy of the University of Vienna.}

\end{document}